\documentclass[aps,onecolumn,showpacs,nofootinbib,showkeys]{revtex4-2}
\usepackage[margin=2.0 cm]{geometry}
\usepackage{feynmp-auto}

\RequirePackage[T1]{fontenc}

\usepackage{epsfig,graphicx,amsmath,amssymb,bm}
\RequirePackage{mathptmx}      
\usepackage{mathrsfs}
\usepackage[english]{babel}
\usepackage{amssymb}
\RequirePackage{color}
\usepackage{changes}
\usepackage{slashed}
\usepackage{multirow}
\usepackage{scalerel}
\usepackage{tikz-feynman}
\usepackage{textcomp}
\usepackage{subcaption}
\usepackage[english]{babel}
\usepackage{float}
\RequirePackage{hyperref}
\hypersetup{
    linktocpage,
    colorlinks,
    citecolor=blue,
    filecolor=black,
    linkcolor=blue,
    urlcolor=blue,
}
\usepackage{nccmath}

\begin{document}

\title{Description of the processes $e^+ e^- \to \pi^+ \pi^-$ and $\tau^- \to \pi^- \pi^0 \nu_\tau$ in the NJL model with value of the vector coupling constant $ g_{\rho} = 6$
}


\author{M.K. Volkov$^{1}$}\email{volkov@theor.jinr.ru}
\author{A.A. Pivovarov$^{1}$}\email{tex$\_$k@mail.ru}
\author{K. Nurlan$^{1,2}$}\email{nurlan@theor.jinr.ru}

\affiliation{$^1$ Bogoliubov Laboratory of Theoretical Physics, JINR, 
                 141980 Dubna, Moscow region, Russia \\
                $^2$ The Institute of Nuclear Physics, Almaty, 050032, Kazakhstan
                }   


\begin{abstract}
It is shown that the processes $e^+ e^- \to \pi^+ \pi^-$ and $\tau^- \to \pi^- \pi^0 \nu_\tau$ can be described in a unified approach in satisfactory agreement with experiment using the vector coupling constant $g_{\rho} = 6$. In this case, in addition to quark loops, it is also necessary to take into account meson loops corresponding to the next order in $1/N_c$. These loops must be taken into account when describing the $\gamma (W) \to \rho$ transition, as well as in interaction of mesons in the final state.


\end{abstract}

\pacs{}

\maketitle


\section{\label{Intro}Introduction}
    The study of the cross section of the $e^+ e^- \to \pi^+ \pi^-$ process is of great interest for a deeper understanding of the internal structure of mesons and determining the main parameters of vector mesons such as mass, width, coupling constant, and etc. In addition, it is important to note that the cross section of the $e^+ e^- \to \pi^+ \pi^-$ process gives more than 70\% contribution to the relation $\sigma\left(e^+ e^- \to \text{hadrons}\right) / \sigma\left(e^+ e^- \to \mu^+ \mu^-\right)$ used to determine the hadronic contribution to the anomalous magnetic moment of the muon \cite{Davier:2010rnx}. The $e^+ e^- \to \pi^+ \pi^-$ process in the energy region < 1 GeV has been studied in many experiments since the 60s of the last century \cite{Auslander:1967xma, Augustin:1969kn, Benaksas:1972ps, Bukin:1978sq, Vasserman:1981xq, Amendolia:1983di, CMD-2:2001ski, KLOE:2004lnj, Achasov:2005rg}. In recent experiments in this energy range, the error has become less than 1\% \cite{BaBar:2009wpw, BaBar:2012bdw, KLOE:2012anl, BESIII:2015equ, SND:2020nwa}. In this energy region, the channel with the intermediate resonance $\rho(770)$ dominates in determining the total cross section.

    The decay widths of $\rho \to e^+ e^-$ and $\rho \to \pi^+ \pi^-$ can be calculated using the vector coupling constant $g_{\rho}$. Current experimental data \cite{ParticleDataGroup:2022pth} allow one to calculate the values of this constant $g_{\rho} = 5$ for the process $\rho \to e^+ e^-$ and $g_{\rho} = 6$ for the process $\rho \to \pi^+ \pi^-$. In theoretical approaches within the vector dominance model, two different values of these constants were used for the pion vector form factor and for the analysis of the $e^+ e^- \to \pi^+ \pi^-$ process cross section (see the work \cite{OConnell:1995nse} and refs therein.) 
    In the work \cite{Volkov:2021fmi}, it was shown that these decays with the inclusion of one-loop meson diagrams corresponding to $1/N_c$ corrections can be described within the Nambu-Jona-Lasinio (NJL) model using a single value $g_ {\rho} = $6.
   
    The NJL quark model \cite{Volkov:1984kq,Volkov:1986zb,Ebert:1985kz,Vogl:1991qt,Klevansky:1992qe,Volkov:1993jw,Ebert:1994mf,Buballa:2003qv,Volkov:2005kw} allows us to describe many internal properties of mesons and the main types of strong, electromagnetic, and weak processes of interaction of mesons at low energies in the leading order in $1/N_c$, where $N_c$ is the number of colors in QCD. 
    Note that when determining the internal parameters of the NJL model, the use of experimental values for the weak pion decay constant $F_{\pi} = 92.4$~MeV ($\pi \to \mu \nu_{\mu}$) and the strong decay constant $g_\rho = 6$ ($\rho \to \pi \pi$) leads to the values of the constituent quark masses $m_u \approx m_d \approx 270$~MeV and the cut-off parameter for the quark loops $\Lambda_q = 1265$~MeV \cite{Volkov:1986zb,Volkov:1993jw,Ebert:1994mf,Volkov:2005kw}. 
    However, the description of the processes $e^+ e^- \to \pi^+ \pi^-$ and $\tau^- \to \pi^- \pi^0 \nu_\tau$ within the NJL model using only quarks loops (in the leading order in $1/N_c$) and a single value $g_\rho = 6$ cannot be consistent with experimental data with sufficient accuracy. 
    In the present work, we show that taking into account meson loops (the next order in $1/N_c$) in the transitions $\gamma (W) \to \rho$ and $\rho \to \pi \pi$ allows one to obtain results in agreement with experimental data for both processes in a unified approach.
   
    In the work \cite{Volkov:2020dvz}, an attempt was made to take into account interactions of pions in the final state to study the processes $e^+ e^- \to \pi^+ \pi^-$ and $\tau^- \to \pi^- \pi ^0 \nu_\tau$ within the NJL model. 
    However, taking into account meson loops also makes a significant contribution to the intermediate state in the $\gamma(W) \to \rho$ transitions which will be shown in the present work.   
   
    In the work \cite{Oller:2000ug}, a unitary approach was developed to account for corrections of final state interactions to tree level amplitudes based on chiral perturbation theory ($\chi$PT) \cite{Gasser:1983yg}. This approach differs significantly from the methods used in the NJL model.

\section{The NJL model Lagrangian in the leading order in $1/N_c$} 
\label{sect:NJL}
	The quark-meson interaction Lagrangian of the NJL model in the leading order in $1/N_c$, which contains the vertices necessary for studying the processes under consideration, takes the following form~\cite{Volkov:1986zb, Volkov:2005kw}:
    \begin{eqnarray}
	\label{Lagrangian}
		\Delta \mathcal{L}_{int} & = &
		\bar{q} \left[ \sum_{j = \pm, 0}\left(\frac{g_{\rho}}{2} \gamma^{\mu} \lambda_{\rho^j} \rho_{\mu}^{j} + i g_{\pi} \gamma^{5} \lambda_{\pi^{j}} \pi^{j}  +
        i g_{K} \gamma^{5} \lambda_{K^{j}} K^{j} \right) + \frac{g_{\omega}}{2} \gamma^{\mu}\lambda_{\omega} \omega_{\mu} + i g_{K} \gamma^{5} \lambda_{\bar{K}^0} \bar{K}^{0}
        \right]q,
	\end{eqnarray}
	where $q$ and $\bar{q}$ are $u$, $d$ and $s$ quark field triplets; $\lambda$ are are linear combinations of the Gell-Mann matrices.

    The quark-meson coupling constants
	\begin{eqnarray}
	\label{Couplings}
		g_{\rho} = g_{\omega} = \left(\frac{2}{3}I_{2}\right)^{-1/2}, \quad g_{\pi} =  \left(\frac{4}{Z_{\pi}}I_{20}\right)^{-1/2},
        \quad g_{K} =  \left(\frac{4}{Z_{K}}I_{11}\right)^{-1/2},
	\end{eqnarray}
	where
	\begin{eqnarray}
	Z_{\pi} = \left[1 - 6\frac{m^{2}}{M_{a_{1}}^{2}}\right]^{-1},
        \quad Z_K = \left[1 - \frac{3(m_s+m_u)^2}{2M_{K_{1A}}^{2}} \right]^{-1},
	\end{eqnarray}
    are additional renormalization constants that arise when taking into account $a_1-\pi$ and $K_1 - K$ transitions; $M_{a_{1}} = 1230 \pm 40$~MeV~\cite{ParticleDataGroup:2022pth} is the mass of the axial-vector meson; $m_{u} \approx m_{d} = m = 270$~MeV and $m_s = 420$~MeV are constituent quark masses arising from spontaneous breaking of chiral symmetry ~\cite{Volkov:2005kw}; $M_{K_{1A}}$ is the effective mass resulting from the mixing of the states $K_1(1270)$ and $K_1(1400)$ 
        \begin{eqnarray}
	M_{K_{1A}} = \left[ \frac{\sin^2(\alpha)}{M^2_{K_1(1270)}} + \frac{\cos^2(\alpha)}{M^2_{K_1(1400)}} \right]^{-1},
	\end{eqnarray}
    where $\alpha=57^{\circ}$ \cite{Suzuki:1993yc, Volkov:2019awd}.

    The integrals appearing from quark loops as a result of the renormalization of the Lagrangian take the following form:
 	\begin{eqnarray}
	\label{int}
		I_{nm} =
		-i\frac{N_{c}}{(2\pi)^{4}}\int\frac{\Theta(\Lambda_q^{2} + k^2)}{(m^2 - k^2)^n(m^2_s - k^2)^m} \mathrm{d}^{4}k 
	\end{eqnarray}
	where $\Lambda_q = 1265$~ MeV is the cutoff parameter and $N_{c} = 3$ is the number of color in QCD. 

\section{The process $e^+ e^- \to \pi^+ \pi^-$} 
\begin{figure}[h]
    \center{\includegraphics[scale = 0.5]{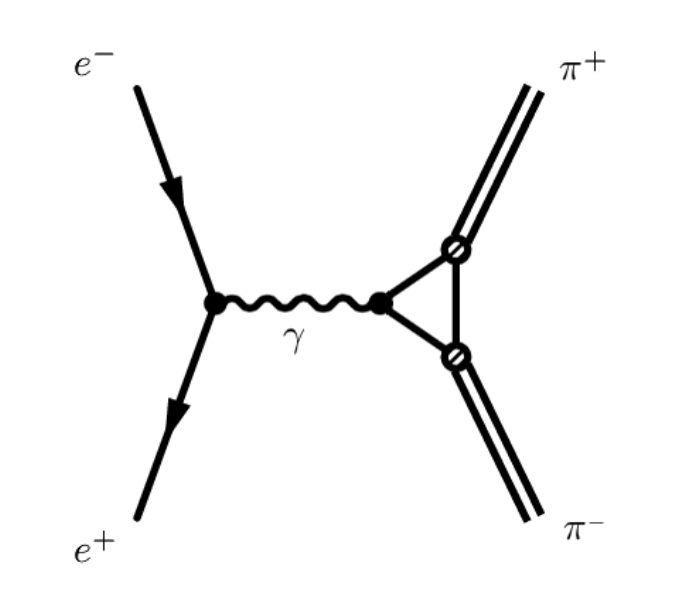}}
    \caption{
    The contact diagram for the process $e^+ e^- \to \pi^+ \pi^-$ in the leading order in $1/N_c$.
    }
    \label{eeContact}
    \end{figure}

\begin{figure}[h]
    \center{\includegraphics[scale = 0.6]{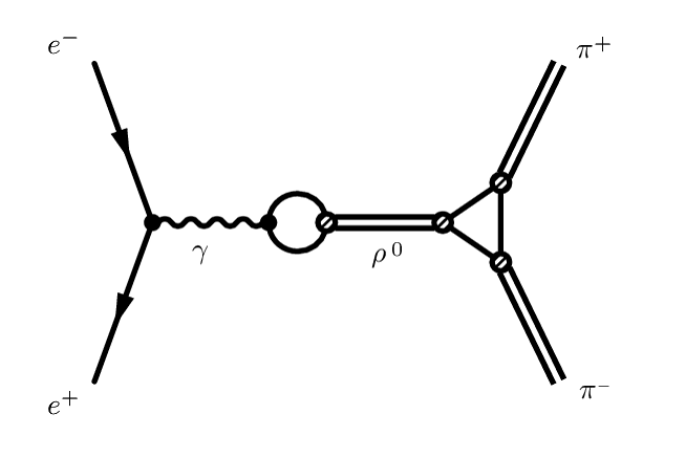}}
    \caption{
    The diagram with the intermediate $\rho$ meson for the process $e^+ e^- \to \pi^+ \pi^-$ in the leading order in $1/N_c$
    }
    \label{eeInterm}
    \end{figure}

\begin{figure}[h]
    \center{\includegraphics[scale = 0.6]{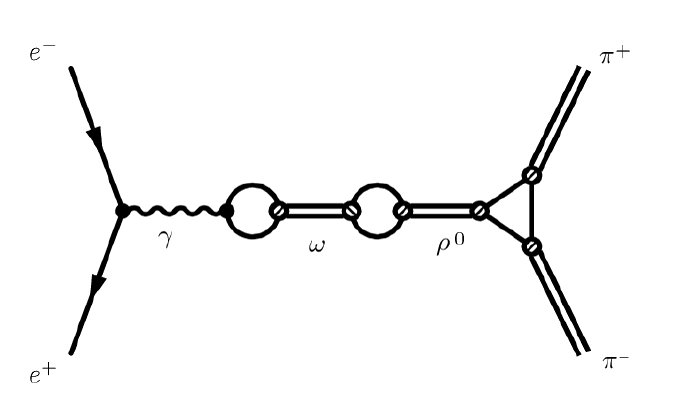}}
    \caption{
    The diagram with the intermediate $\omega$ meson for the process $e^+ e^- \to \pi^+ \pi^-$ in the leading order in $1/N_c$
    }
    \label{eeOmega}
    \end{figure}

    The diagrams describing the process $e^+ e^- \to \pi^+ \pi^-$ in the leading order in $1/N_c$ are presented in Figs. \ref{eeContact}, \ref{eeInterm} and \ref{eeOmega}.
    
    In order to take into account $1/N_c$ corrections when calculating this process, it is necessary to consider an additional transition of the photon to the intermediate $\rho$-meson through the meson loop and the interaction of pions in the final state.

    The transition between the photon and the $\rho^0$-meson through meson loops within the NJL model was obtained in \cite{Volkov:2021fmi}:
    \begin{eqnarray}
		-2\frac{\sqrt{\pi\alpha_{em}}}{6q^2} g_{\rho} \left(2 I_{\rho\pi} + I_{\rho K}\right),
	\end{eqnarray}
    where
    \begin{eqnarray}
    \label{gamma_rho}
		I_{\rho\pi} & = & \frac{1}{16\pi^2} \left[2\Lambda_M^2 + \left(q^2 - 6M_{\pi}^2\right)\ln\left(1 + \frac{\Lambda_M^2}{M_{\pi}^2}\right) + 2\left(2\Lambda_M^2 + q^2 - 4M_{\pi}^2\right) D\left(\Lambda_M^2\right) \arctan\frac{1}{D\left(\Lambda_M^2\right)} \right.\nonumber\\
        && \left. + 2q^2 D(0)^3\arctan\frac{1}{D(0)}\right] - \frac{3}{8\pi^2} \left[\Lambda_M^2 - M_{\pi}^2 \ln\left(\frac{\Lambda_M^2}{M_{\pi}^2} + 1\right)\right],        
	\end{eqnarray}
    \begin{eqnarray}
        D\left(\Lambda_M^2\right) & = & \sqrt{4\frac{\Lambda_M^2 + M_{\pi}^2}{q^2} - 1},
    \end{eqnarray}
    where $q$ is the $W$ boson momentum; $I_{\rho\pi}$ and $I_{\rho K}$ correspond to $\pi$ and K mesons loops; $I_{\rho K}$ is obtained from $I_{\rho\pi}$ with the replacement of $M_\pi \to M_K$. 
    
\begin{figure}[h]
    \center{\includegraphics[scale = 0.3]{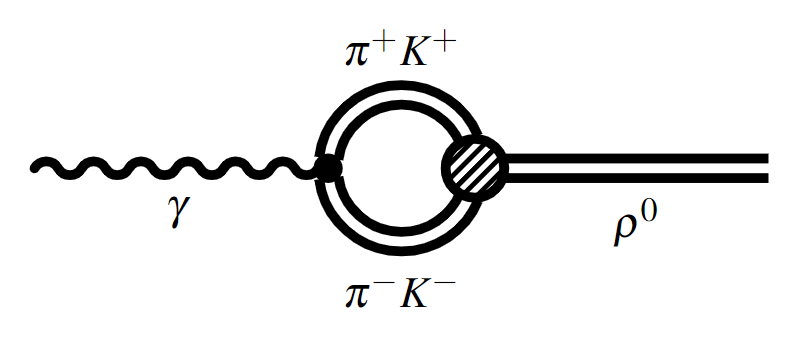}}
    \caption{
    The diagram describing the meson loop with two vertices for the transition $\gamma \to \rho^0$
    } 
    \label{gamma_rho_1}
    \end{figure}

\begin{figure}[h]
    \center{\includegraphics[scale = 0.3]{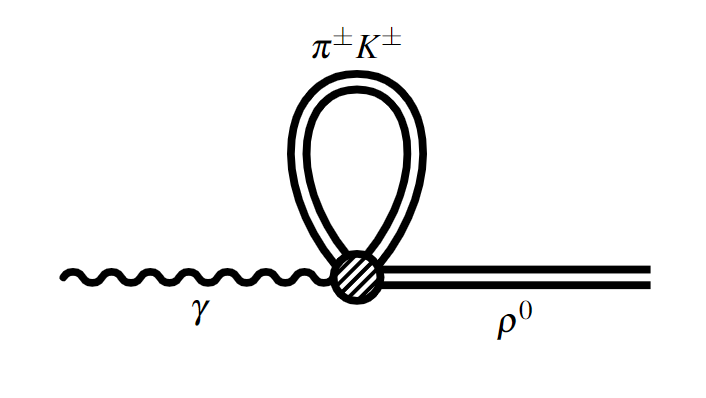}}
    \caption{
    The tadpole diagram describing the transition $\gamma \to \rho^0$
    } 
    \label{gamma_rho_2}
    \end{figure}

    The first term in (\ref{gamma_rho}) corresponds to the loop with two-vertices presented in Fig. \ref{gamma_rho_1} and the second term corresponds to the tadpole diagram presented in Fig. \ref{gamma_rho_2}.

   %
   %
   %

  %
  %
  %

    The final state interaction can be taken into account by considering the meson triangle in the $\rho^0 \to \pi^+ \pi^-$ decay. This interaction was studied in the paper \cite{Volkov:2020dvz}. The corresponding corrections to the tree level-amplitude take the form:
    \begin{eqnarray}
		i g_{\rho}^3 \left[\frac{I_{1M}}{M_{\rho}^2} + I_{2M}\right] \left(p_{+} - p_{-}\right)^{\mu},
	\end{eqnarray}
    where
    \begin{eqnarray}
		I_{1M} & = & \frac{-i}{(2\pi)^4} \int \frac{\Theta\left(\Lambda_M^2 + k^2\right)}{M_{\rho}^2 - k^2} d^4k = \frac{1}{(4\pi)^2} \left[\Lambda_M^2 - M_{\rho}^2 \ln\left(\frac{\Lambda_M^2}{M_{\rho}^2} + 1\right)\right],\nonumber\\
        I_{2M} & = & \frac{-i}{(2\pi)^4} \int \frac{\Theta\left(\Lambda_M^2 + k^2\right)}{(M_{\rho}^2 - k^2)(M_{\pi}^2 - k^2)} d^4k = \frac{1}{(4\pi)^2} \frac{1}{M_{\rho}^2 - M_{\pi}^2} \left[ M_{\rho}^2 \ln\left(\frac{\Lambda_M^2}{M_{\rho}^2} + 1\right) -  M_{\pi}^2 \ln\left(\frac{\Lambda_M^2}{M_{\pi}^2} + 1\right)\right],
	\end{eqnarray}
    $\Lambda_M$ is the meson loop cutoff parameter. 
    
    The amplitude of the process $e^+ e^- \to \pi^+ \pi^-$ with corrections from meson loops takes the following form:
    \begin{eqnarray}
		M & = & -\frac{4\pi\alpha_{em}}{q^2} \left[\left(T_c + T_{\rho} + T_{\rho\omega}\right)\left(1 +  T_{\triangle}\right) + T_{\circ} \right]l_{\mu}\left(p_{+} - p_{-}\right)^{\mu},
	\end{eqnarray}
    where $q$ is the momentum of the intermediate photon; $T_c$, $T_{\rho}$ and $T_{\rho\omega}$ are the contributions from the contact diagram and diagrams with the intermediate $\rho$ and $\omega$ mesons
    \begin{eqnarray}
		&T_c = 1,& \nonumber\\
        &T_{\rho} = \frac{q^2}{M_{\rho}^2 - q^2 - i\sqrt{q^2}\Gamma_{\rho}},& \nonumber\\
        &T_{\rho\omega} = \frac{g_{\rho}^2}{9}\frac{q^4\left(I_{20}(u) - I_{20}(d)\right)}{\left(M_{\rho}^2 - q^2 - i\sqrt{q^2}\Gamma_{\rho}\right)\left(M_{\omega}^2 - q^2 - i\sqrt{q^2}\Gamma_{\omega}\right)},& \nonumber
	\end{eqnarray}
    where the factor
    \begin{eqnarray}
		T_{\triangle} = g_{\rho}^2\left(\frac{I_{1M}}{M_{\rho}^2} + I_{2M}\right)
	\end{eqnarray}
    corresponds to the correction from final state interaction.  
    The term $T_{\circ}$ describes the correction from the meson loop in the intermediate state in the $\gamma \to \rho$ transition
    \begin{eqnarray}
		T_{\circ} =\frac{g_{\rho}^2}{6} \left(2 I_{\rho\pi} + I_{\rho K}\right) \frac{q^2}{M_{\rho}^2 - q^2 - i\sqrt{q^2}\Gamma_{\rho}}.
	\end{eqnarray}
    
    For the width of the intermediate $\rho$ meson, we will use a function of energy in the form $\Gamma_\rho \to \Gamma_\rho(q^2)$
    \begin{eqnarray}
		\Gamma_\rho(q^2) = \frac{g^2_\rho}{48 \pi q^2} {\left( q^2 - 4 M^2_\pi \right)}^{3/2}.
	\end{eqnarray}
    
\begin{figure}[h]
    \center{\includegraphics[scale = 0.7]{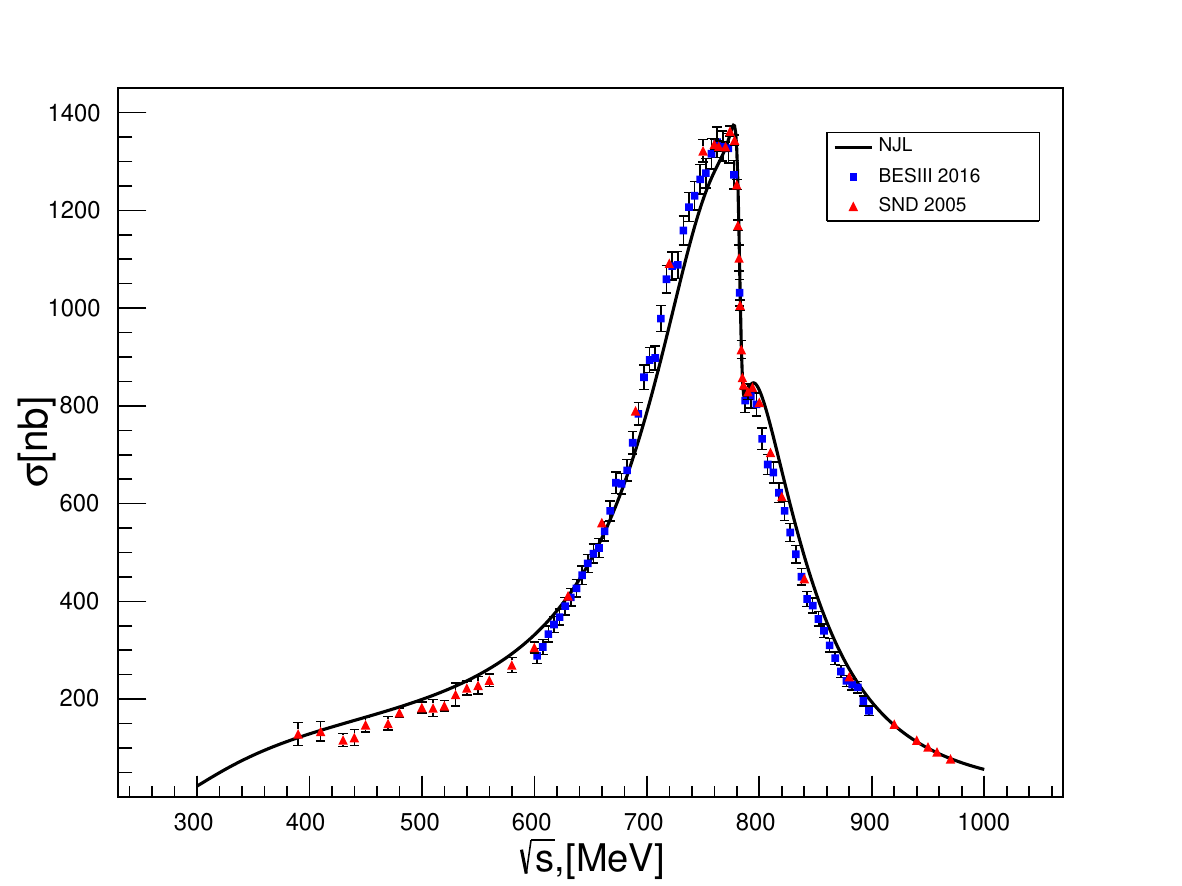}}
    \caption{
    The comparison of the total cross section obtained in the framework of the NJL model for the process $e^+e^- \to \pi^+\pi^-$ with the experimental data of the SND \cite{Achasov:2005rg} and BES III \cite{BESIII:2015equ} collaborations
    }
    \label{crosssection}
    \end{figure}

    The integrals $I_{20}(u)$ and $I_{20}(d)$ are given in (\ref{int}) with the replacement of $mu$ with $m_d$. The difference in the masses of these quarks is taken as 4 MeV \cite{Volkov:1986zb,Volkov:2020dvz}. 
    The contribution of the term $T_{\rho\omega}$ practically does not affect the height of the cross-section dependence graph due to the small value of the difference $I_2(u) - I_2(d)$ and the factor $1/9$ in the amplitude. At the same time, it affects its shape.
    We fix the value of the cutoff parameter for meson loops equal to $\Lambda_{M} = 650$~MeV based on the known experimental data of the cross section (Figure~\ref{crosssection})$\Lambda_{M} = 650$~MeV.

\section{The process $\tau^- \to \pi^- \pi^0 \nu_\tau$} 

    The hadron current obtained in the process $e^+e^- \to \pi^+ \pi^-$ is associated with the decay $\tau^- \to \pi^- \pi^0 \nu_\tau$ by the isospin transformation in the framework of the Conserved Vector Current theorem (CVC) \cite{Gilman:1977cc, CLEO:1999dln}. The cross section of the process $e^+e^- \to \pi^+ \pi^-$ can be represented through the spectral hadron function $v^{\pi\pi}(s)$.
    \begin{eqnarray}
		\sigma(e^+e^- \to \pi^+ \pi^-) =\left( \frac{4 \pi^2 \alpha^2_{em}}{s} \right) v^{\pi\pi}(s).
	\end{eqnarray}
    
    The appropriate differential width of the decay $\tau^- \to \pi^- \pi^0 \nu_\tau$ takes the form
    \begin{eqnarray}
		\frac{d\Gamma(\tau^- \to \pi^- \pi^0 \nu_\tau)}{dq^2} = \frac{G^2_F V^2_{ud} S^{\pi\pi}_{EW}}{32 \pi^2 M^3_\tau} 
  {\left( M^2_\tau - q^2 \right)}^2 \left( M^2_\tau + 2q^2 \right) v^{\pi\pi^0}(q^2),
	\end{eqnarray}
    where in the framework of CVC $v^{\pi\pi}(s)=v^{\pi\pi^0}(q^2)$, $s=q^2$.

    Defining the spectral hadron function for the process $e^+e^- \to \pi^+ \pi^-$ within the CVC, one can obtain the estimation for the branching fractions of the decay $\tau^- \to \pi^- \pi^0 \nu_\tau$
    \begin{eqnarray}
    \label{br1}
        Br(\tau^- \to \pi^- \pi^0 \nu_\tau) = (26.2 \pm 1.3)\%.
	\end{eqnarray}

    The performed calculations show that the considered processes have the main contribution from the vector channel with the $\rho$ meson. In the $1/N_c$ corrections, the main contribution is given by the pion loops. Therefore, the uncertainty in the $SU(2) \times SU(2)$ model is estimated as 5\% \cite{Volkov:2017arr}. 
    This result is in the agreement with the experimental data~\cite{ParticleDataGroup:2022pth}:
    \begin{eqnarray}
		Br(\tau^- \to \pi^- \pi^0 \nu_\tau)_{exp} = (25.5 \pm 0.1)\%
	\end{eqnarray}

\begin{figure}[h]
    \center{\includegraphics[scale = 0.8]{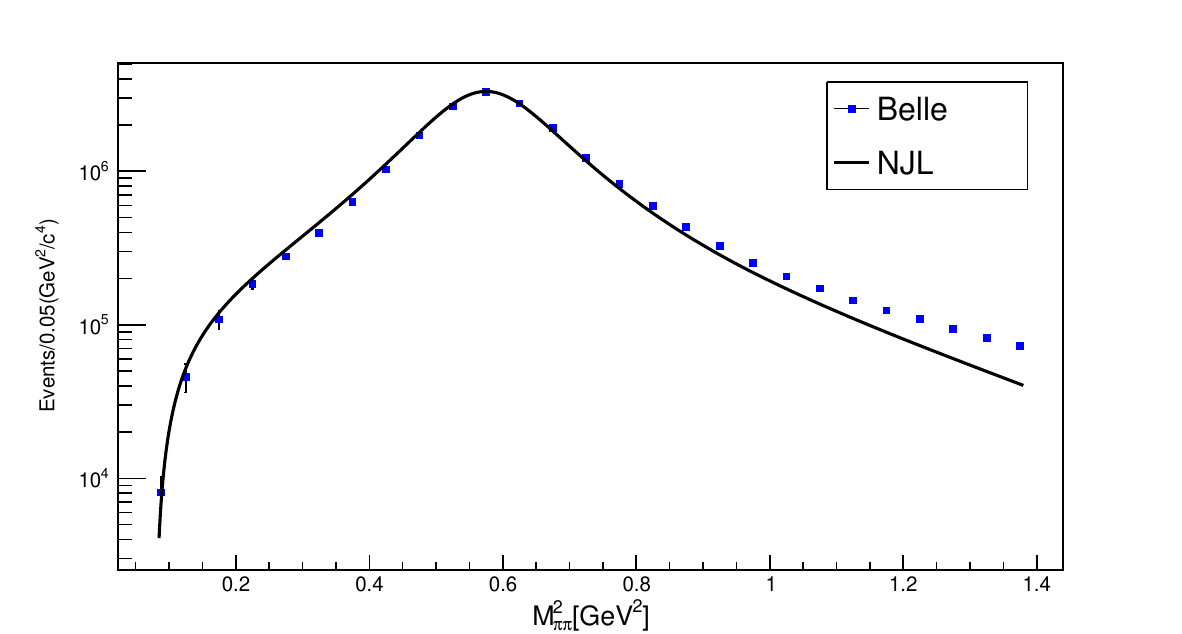}}
    \caption{
    The calculated invariant mass $M_{\pi\pi^0}$ distribution for the process $\tau \to \pi^- \pi^0 \nu_\tau$. The experimental data taken from \cite{Belle:2008xpe}
    } 
    \label{massdistr}
    \end{figure}

    The use of the spectral hadronic function within the CVC is possible due to the presence of only a vector channel in both processes. 
    It should be note that by using the spectral hadron function for the decay $\tau^- \to \pi^- \pi^0 \nu_\tau$, we do not take into account the contribution $T_{\rho\omega}$ in the amplitude due to the absence of the $\omega$ meson in the intermediate state.
    The calculated invariant mass $M_{\pi\pi^0}$ distribution for the $\tau$ decay and its comparison with the experimental data are shown in Fig. \ref{massdistr}.

\begin{figure}[h]
    \center{\includegraphics[scale = 0.4]{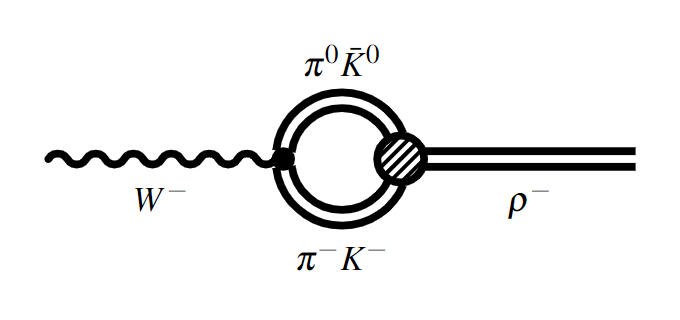}}
    \caption{
    The two-vertices meson loop of the transition $W^- \to \rho^-$
    } 
    \label{W_rho_1}
    \end{figure}

\begin{figure}[h]
    \center{\includegraphics[scale = 0.4]{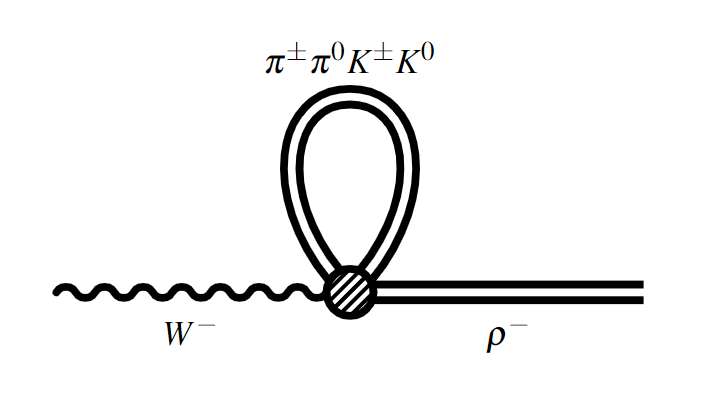}}
    \caption{
    The tadpole diagram describing the transition  $W^- \to \rho^-$
    } 
    \label{W_rho_2}
    \end{figure}

    It is important to note that the estimation for the branching fraction of $\tau^- \to \pi^- \pi^0 \nu_\tau$ can be obtained by direct calculation as it has been done for the process $e^+ e^- \to \pi^+ \pi^-$.
    The one-loop meson correction in the final state in this process is described similarly to the process $e^+ e^- \to \pi^+ \pi^-$. To take into account the correction in the intermediate state, it is necessary to obtain the transition $W^- \to \rho^-$ through the meson loops.
    This transition differs from the transition $\gamma \to \rho^0$ by the presence of additional diagrams (Figures \ref{W_rho_1} and \ref{W_rho_2}). However, the calculations show that the structure of this transition is similar to the case $\gamma \to \rho^0$ and differs only by the numeric coefficient:    
    \begin{eqnarray}
		- 2^{1/4} \frac{M_{W}G_{F}V_{ud}}{6q^2} g_{\rho} \left(2 I_{\rho\pi} + I_{\rho K}\right).
	\end{eqnarray}

    %
     %
      %

   %
   %
      %

    As a result, the amplitude of the process $\tau^- \to \pi^- \pi^0 \nu_\tau$ taking into account the $1/N_c$ corrections mentioned above takes the form
    \begin{eqnarray}
		M & = & -G_{F}V_{ud} \left\{\left[T_c + T_{\rho}\right]\left[1 + T_{\triangle}\right] + T_{\circ} \right\}L_{\mu}\left(p_{-} - p_{0}\right)^{\mu},
	\end{eqnarray}
    where $L_{\mu}$ is the week lepton current.

    As one can see, the structure of this amplitude is mostly similar to the structure of the amplitude of the process $e^+ e^- \to \pi^+ \pi^-$. However, this amplitude does not include the contribution from the $\omega$ meson. And taking into account the one-loop $1/N_c$ corrections has not led to the break of CVC.

    The value of the cut-off parameter for the meson loops $\Lambda_{M} = 650$~MeV obtained above is used for the calculation of the branching fractions   
    \begin{eqnarray}
		Br(\tau^- \to \pi^- \pi^0 \nu_\tau) = (26.6 \pm 1.3)\%
	\end{eqnarray}

    This result in the framework of the model errors is in agreement with the result (\ref{br}) obtained above by using the spectral function within CVC (\ref{br1}).

\section{Conclusion} 
    The calculations in the framework of the NJL model show that two important processes $e^+ e^- \to \pi^+ \pi^-$ and $\tau^- \to \pi^- \pi^0 \nu_\tau$ can be described by using the single value of the vector coupling constant $g_{\rho} = 6$. 
    For this purpose, both contributions from the quark loops and additional corrections from the meson loops corresponding to the next-to-leading order in the $1/N_c$ expansion were considered. Indeed, direct calculations of the process $e^+ e^- \to \pi^+ \pi^-$ cross section and the width of the decay $\tau^- \to \pi^- \pi^0 \nu_\tau$ by using the equal value of the constant $g_{\rho}$ in the transitions $\gamma \to \rho^0$ and $W^- \to \rho^-$ and in the decays $\rho^0 \to \pi^+ \pi^-$ and $\rho^- \to \pi^- pi^0$ lead to consistent results in satisfactory agreement with the experimental data, wherein the single cut-off parameter for the meson loops is used $\Lambda_{M} = 650$~MeV.

    The width of the decay $\tau^- \to \pi^- \pi^0 \nu_\tau$ can also be obtained from the process $e^+ e^- \to \pi^+ \pi^-$ using the Conserved Vector Current theorem. The use of CVC becomes possible because the decay $\tau^- \to \pi^- \pi^0 \nu_\tau$ is fully determined by the vector channel. It is necessary to note that this fact differs this process from the other $\tau$ decay modes where the main contributions are given by the axial vector channels \cite{Volkov:2017arr} and such transformation by using CVC is impossible.

    The model predictions for the distribution of invariant masses $M_{\pi\pi}$ and the comparison with the experimental data \cite{CLEO:1999dln} are represented in Fig. \ref{massdistr}. In the energy range above 1~GeV the theoretical data diverge from the experiment. In our opinion, this is because in the present work the contributions from the excited meson states and mainly from the meson $\rho(1450)$ are not taken into account \cite{Belle:2008xpe,CLEO:1999dln}.

\subsection*{Acknowledgments}
    The authors are grateful to A.~B.~Arbuzov for useful discussions.


\end{document}